\documentclass[11pt]{article}
\usepackage[a4paper,margin=2.2cm]{geometry}
\usepackage{amsmath,amssymb,bm}
\usepackage{graphicx}
\usepackage{booktabs}
\usepackage{cite}
\usepackage{hyperref}
\usepackage{microtype}
\hypersetup{hidelinks}
\graphicspath{{figures/}}
\allowdisplaybreaks

\title{Conservative Orbital Resonance Networks as Local Environmental Diagnostics for Extreme Mass Ratio Inspirals}
\author{%
\normalsize Mona Kumari$^{1}$, Anirudh Pradhan$^{2}$, K. Ghaderi$^{3,*}$, and Javlon Rayimbaev$^{4,5,6}$\\[0.45em]
\small $^{1}$Department of CEA, GLA University, Mathura - 281406, Uttar Pradesh, India\\
\small $^{2}$Centre for Cosmology, Astrophysics and Space Science (CCASS), GLA University, Mathura 281406, U.P., India\\
\small $^{3}$Department of Physics, Mari.C., Islamic Azad University, Marivan, Iran\\
\small $^{4}$Harbin Institute of Technology, Harbin 150001, People's Republic of China\\
\small $^{5}$University of Tashkent for Applied Sciences, Str. Gavhar 1, Tashkent 100149, Uzbekistan\\
\small $^{6}$Institute of Theoretical Physics, National University of Uzbekistan, Tashkent 100174, Uzbekistan\\
\small $^{*}$Corresponding author: \texttt{k.ghaderi@iau.ac.ir}
}
\date{}

\begin{document}
\maketitle

\begin{abstract}
Low order orbital resonances provide localized probes of the conservative geometry sampled by extreme mass ratio inspirals. We study fixed eccentricity and inclination slices of the ordered $3{:}2$, $5{:}3$, and $2{:}1$ polar to radial resonance surfaces in source consistent anisotropic black hole environments, treating rotation strictly to first order. A horizon regularized Dehnen sequence produces unequal resonance displacements, reaching $-7.17\%$, $-5.45\%$, and $-3.94\%$ in a fixed global benchmark. After matching both the total environmental mass and the enclosed mass at $10M_\bullet$, we compare Dehnen and Einasto inspired families. Two dimensionless network spacing ratios are organized primarily by the local logarithmic enclosed mass slope: for the common $\beta=2$ regularization their maximum cross family departure from a shared linear relation is $4.1\times10^{-4}$, while varying the horizon taper through $\beta=2$--$4$ increases the departure to at most $2.1\times10^{-3}$. The unequal ordering of the three resonance displacements holds at every point of a directly evaluated finite eccentricity and inclination grid. A six observable conservative response analysis is numerically full rank but strongly ill conditioned and nearly aligned with orbital nuisance directions. The result therefore establishes local conservative profile sensitivity, not unique tomography or a detector level parameter recovery forecast.
\end{abstract}


\section{Introduction}
\label{sec:intro}

Extreme mass ratio inspirals (EMRIs) are precision probes of massive black holes and their surroundings because a compact secondary executes many relativistic orbital cycles while slowly evolving under radiation reaction \cite{Colpi2024,BarackPound2019}. Generic bound motion around a rotating black hole contains distinct radial, polar, and azimuthal frequencies, and the geometry of low order orbital resonances has been mapped across relativistic black hole parameter space \cite{BrinkGeyerHinderer2015}. During an inspiral the radial and polar frequencies can become commensurate,
\begin{equation}
 n\Omega_r+k\Omega_\theta=0,
 \qquad n,k\in\mathbb Z,
 \label{eq:res_condition_intro}
\end{equation}
so that the adiabatic approximation requires special care near a transient resonance \cite{FlanaganHinderer2012,RuangsriHughes2014,SperiGair2021}. Resonance induced changes depend on the resonant forcing and entrance phase, and recent analyses have reinforced that their impact cannot be represented by a universal jump independent of the orbital state \cite{FlanaganHughesRuangsri2014,LevatiCardenas2026}.

Astrophysical EMRIs need not evolve in vacuum. Surrounding dark matter or baryonic structures can perturb both conservative geodesic dynamics and dissipative evolution. Dark matter induced drag and waveform modifications have been studied in several regimes \cite{HannukselaEtAl2020,ZhangFuDai2024,DuqueEtAl2024,ZhaoDaiGong2025}, while relativistic treatments have begun to incorporate external gravitating matter directly into EMRI dynamics \cite{PolcarEtAl2022} and, more recently, into a two parameter framework for gravitational radiation in matter environments \cite{PolcarWitzany2025}. Source consistent black hole solutions supported by anisotropic matter provide a complementary setting in which the environmental geometry is connected to an explicit stress tensor \cite{CardosoEtAl2022,FigueiredoEtAl2023,FernandesCardoso2025}. Slow rotation further introduces matter dependent frame dragging. Recent work has propagated this effect into circular orbit quantities, radial and vertical epicyclic frequencies, and epicyclic resonance positions \cite{DattaSingha2026}. The calculation below addresses a distinct regime and question: finite eccentricity polar to radial fundamental frequency resonances are treated as a correlated multi resonance geometry, rather than as isolated circular epicyclic resonances, and are subjected to matched mass cross family tests and an explicit nuisance conditioning analysis.

The central question considered here is deliberately narrower than full environmental waveform inference. Different low order commensurabilities occur at different radii, so a correlated set of resonance locations samples the radial gravitational field at several strong field scales. We ask which local information about the matter distribution survives after the overall environmental mass amplitude is matched, whether that information persists across distinct profile families and regularizations, and how strongly it is degenerate with ordinary orbital parameters. Throughout the benchmark comparison, the resonance loci are evaluated on a fixed $(e,\iota)$ slice of the conservative resonance surfaces. They are therefore not identified with the sequential crossing radii of one dissipative inspiral, for which $e$, $\iota$, and the actions evolve between crossings. The conservative locations are useful precisely because their definition does not require an environmental flux model or a prescription for resonance jumps.

The contribution is threefold. First, we recompute finite eccentricity resonance geometry with a perturbatively consistent first order expansion in the rotational parameter and explicitly follow the outer librating resonance branch. Second, we compare a horizon regularized Dehnen family with an independent Einasto inspired family while matching both the total environmental mass and the enclosed mass at a reference radius, and then vary the horizon taper to test sensitivity to the regularization. The resulting network shape coordinates are organized primarily, but not completely, by the local logarithmic enclosed mass slope. Third, we quantify degeneracy with eccentricity, inclination, rotation amplitude, and matter rotation using a conservative response matrix. The matrix is numerically full rank but strongly ill conditioned, which rules out robust local inversion from these conservative observables alone.

No timing or accumulated phase observable is used for parameter inference in this work. Peters type transport is derived in the weak field and has not been shown to provide a controlled subpercent systematic error for the present environmental strong field background; moreover, sequential resonance clocks are altered by resonance jumps and environmental drag \cite{StratenyEtAl2026}. These effects define the boundary between the conservative result established here and a future waveform level analysis.

The paper is organized as follows. Section~\ref{sec:geometry} constructs the source consistent static and slowly rotating backgrounds. Section~\ref{sec:orbits} derives the bound motion and strict first order resonance calculation. Section~\ref{sec:network} presents the Dehnen resonance network and matched local mass test. Section~\ref{sec:crossfamily} introduces the independent profile comparison and conservative nuisance analysis. Section~\ref{sec:scope} connects the conservative construction to dissipative inspiral observables and states the required additional ingredients. Section~\ref{sec:numerics} gives numerical verification and limiting cases, and Sec.~\ref{sec:conclusion} summarizes the conclusions.

\section{Source Consistent Environmental Geometry}
\label{sec:geometry}

\subsection{Static anisotropic background}

We use geometrized units $G=c=1$. The static line element is
\begin{equation}
 ds_0^2=-A(r)dt^2+\frac{dr^2}{B(r)}+r^2(d\theta^2+\sin^2\theta\,d\phi^2),
 \qquad
 B(r)=1-\frac{2m(r)}{r}.
 \label{eq:metric_static}
\end{equation}
The effective matter source is an Einstein cluster type anisotropic stress tensor,
\begin{equation}
 T^\mu{}_{\nu}=\mathrm{diag}[-\rho,0,p_t,p_t].
 \label{eq:stress}
\end{equation}
The independent Einstein equations reduce to
\begin{align}
 m'(r)&=4\pi r^2\rho(r),
 \label{eq:mass_eq}\\
 \frac{A'}{A}&=\frac{2m}{r(r-2m)},
 \label{eq:lapse_eq}\\
 p_t&=\frac{m\rho}{2(r-2m)}.
 \label{eq:pt_eq}
\end{align}
Asymptotic flatness fixes
\begin{equation}
 A(r)=\exp\left[-2\int_r^\infty\frac{m(s)}{s[s-2m(s)]}\,ds\right].
 \label{eq:A_quad}
\end{equation}
For $m(r)=M_\bullet$, Eq.~\eqref{eq:A_quad} gives $A=B=1-2M_\bullet/r$, providing the vacuum Schwarzschild limit.

The vanishing radial pressure makes this construction useful as an effective collisionless environment, but it should not be interpreted as a complete microscopic matter model at all radii. In particular, Einstein cluster profiles with $p_r=0$ generically violate the dominant energy condition sufficiently close to the horizon and do not supply a causal microscopic sound speed prescription \cite{Datta2024}. We therefore monitor energy conditions over the orbital domain used below and assign no fluid sound speed interpretation to the effective source.

\subsection{Dehnen and Einasto inspired profile families}

The first family is
\begin{equation}
 m_D(r)=M_\bullet+M_{\rm h}F_D(r),
 \label{eq:mass_dehnen}
\end{equation}
with
\begin{equation}
 F_D(r)=\left(\frac{r}{r+a_0}\right)^{3-\gamma}
 \left(1-\frac{2M_\bullet}{r}\right)^\beta,
 \qquad 0\leq\gamma<3,\quad\beta\geq2.
 \label{eq:F_dehnen}
\end{equation}
Away from the horizon this is motivated by the Dehnen family, which includes the Hernquist profile as a special case \cite{Hernquist1990,Dehnen1993}. The taper enforces $m_D(2M_\bullet)=M_\bullet$ and leaves $m_D(\infty)=M_\bullet+M_{\rm h}\equiv M_{\rm ADM}$. Its density is
\begin{equation}
 \rho_D(r)=\frac{M_{\rm h}F_D(r)}{4\pi r^3}
 \left[(3-\gamma)\frac{a_0}{r+a_0}
 +\frac{2\beta M_\bullet}{r-2M_\bullet}\right].
 \label{eq:rho_dehnen}
\end{equation}
For $\beta\geq2$, $\rho_D={\cal O}[(r-2M_\bullet)^{\beta-1}]$ and $p_t={\cal O}[(r-2M_\bullet)^{\beta-2}]$, so the stress eigenvalues remain finite at the horizon. The choice $(\gamma,\beta)=(1,2)$ reduces to the relativistic Hernquist mass profile introduced in Ref.~\cite{CardosoEtAl2022}.

To test whether the resonance response is tied to one profile ansatz, we introduce a second, independent family based on the Einasto form used in dark matter halo studies \cite{NavarroEtAl2004,GrahamEtAl2006}. Its normalized cumulative mass fraction is written
\begin{equation}
 P_E(r)=P\left(\frac{3}{\alpha_E},\frac{2}{\alpha_E}
 \left(\frac{r}{r_{-2}}\right)^{\alpha_E}\right),
 \label{eq:einasto_P}
\end{equation}
where $P(a,x)$ is the regularized lower incomplete gamma function. We define the horizon regularized cumulative fraction
\begin{equation}
 F_E(r)=P_E(r)\left(1-\frac{2M_\bullet}{r}\right)^\beta,
 \qquad
 m_E(r)=M_\bullet+M_{\rm h}F_E(r).
 \label{eq:einasto_mass}
\end{equation}
Because of the relativistic taper, Eq.~\eqref{eq:einasto_mass} is an Einasto inspired environmental family rather than an unmodified Newtonian Einasto halo. The symbol $r_{-2}$ denotes the scale radius inherited from the untapered Einasto profile; after multiplication by the horizon factor it need not coincide exactly with the radius where the final density has logarithmic slope $-2$. The family nevertheless has the same total environmental mass and supplies a structurally different radial mass distribution for the cross family test.

We define
\begin{equation}
 \mu=\frac{M_{\rm h}}{M_\bullet},\qquad
 \alpha=\frac{a_0}{M_\bullet},\qquad
 \epsilon_{10}=\frac{m(10M_\bullet)-M_\bullet}{M_\bullet},
 \label{eq:env_coords}
\end{equation}
and the local logarithmic enclosed mass slope
\begin{equation}
 \kappa_{10}=\left.\frac{d\ln[m(r)-M_\bullet]}{d\ln r}\right|_{r=10M_\bullet}.
 \label{eq:kappa10}
\end{equation}
For the Dehnen family,
\begin{equation}
 \kappa_{10}=(3-\gamma)\frac{\alpha}{10+\alpha}+\frac{2\beta}{8}.
 \label{eq:kappa_dehnen}
\end{equation}
These two local coordinates separate the amplitude of the enclosed environmental mass from its radial gradient.

\subsection{Linear rotation}

\begin{figure}[t]
\centering
\begin{minipage}[t]{0.32\linewidth}\centering
\includegraphics[width=\linewidth]{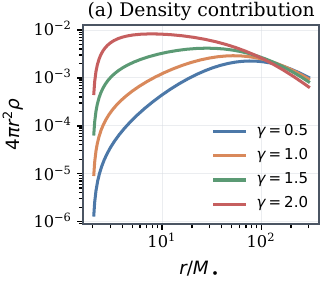}
\end{minipage}\hfill
\begin{minipage}[t]{0.32\linewidth}\centering
\includegraphics[width=\linewidth]{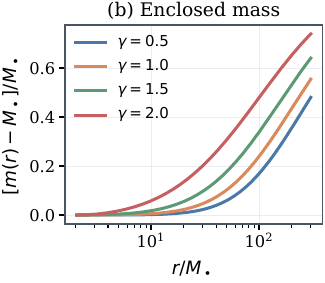}
\end{minipage}\hfill
\begin{minipage}[t]{0.32\linewidth}\centering
\includegraphics[width=\linewidth]{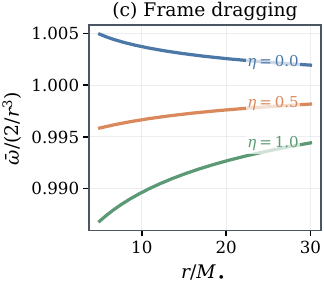}
\end{minipage}
\caption{Source consistent environmental geometry. Panels (a) and (b) show the fixed global Dehnen sequence $(\mu,\alpha,\beta)=(1,100,2)$. Panel (c) shows the normalized frame dragging response for the baseline $\gamma=1.5$ environment while varying the matter rotation parameter.}
\label{fig:geometry}
\end{figure}

Rotation is introduced at first order in an asymptotic spin parameter through a Hartle type expansion \cite{Hartle1967},
\begin{align}
 ds^2={}&-A\,dt^2+B^{-1}dr^2+r^2d\theta^2+r^2\sin^2\theta\,d\phi^2
 \nonumber\\
 &-2r^2\sin^2\theta\,\omega(r)dt\,d\phi+{\cal O}(\chi_\infty^2).
 \label{eq:slow_metric}
\end{align}
We use the effective matter rotation law
\begin{equation}
 \Omega_{\rm m}(r)=\eta\omega(r),\qquad 0\leq\eta\leq1,
 \label{eq:eta}
\end{equation}
for which the $t\phi$ Einstein equation is \cite{DattaSingha2026}
\begin{align}
0={}&\omega''+
\left[4-\frac{4\pi r^3\rho}{r-2m}\right]\frac{\omega'}{r}
\nonumber\\
&+\frac{8\pi r(3m-2r)\rho}{(r-2m)^2}(\omega-\Omega_{\rm m}).
\label{eq:omega}
\end{align}
At large radius,
\begin{equation}
 \omega(r)=\frac{2J_\infty}{r^3}+{\cal O}(r^{-4}),
 \qquad
 \chi_\infty=\frac{J_\infty}{M_\bullet^2}.
 \label{eq:spin}
\end{equation}
In matter, $J_\infty$ is the asymptotic angular momentum normalization of the combined slowly rotating configuration; accordingly, $\chi_\infty$ is used here as an asymptotic rotation amplitude and should not be identified uniquely with the central black hole spin. In the vacuum limit it reduces to the usual Kerr normalization at first order. The closure in Eq.~\eqref{eq:eta} interpolates between a static medium ($\eta=0$) and local corotation with the frame dragging field ($\eta=1$); intermediate values are a controlled phenomenological interpolation rather than a microphysical rotation law. The calculations below use $M_\bullet=1$, $\omega=\chi_\infty\bar\omega$, and the baseline values in Table~\ref{tab:baseline}. Figure~\ref{fig:geometry} summarizes the corresponding density, enclosed mass, and frame dragging structure.

\begin{table}[t]
\centering
\caption{Baseline structural parameters. They define a controlled strong field benchmark, not an astrophysical population prior.}
\label{tab:baseline}
\begin{tabular}{lll}
\toprule
Parameter & Baseline & Meaning\\
\midrule
$\mu$ & $1$ & total environmental mass ratio\\
$\alpha$ & $100$ & Dehnen radial scale\\
$\beta$ & $2$ & horizon taper exponent\\
$\eta$ & $0.5$ & matter rotation interpolation\\
$\chi_\infty$ & $0.05$ & asymptotic spin amplitude\\
$e$ & $0.2$ & orbital eccentricity\\
$\iota$ & $45^\circ$ & orbital inclination\\
\bottomrule
\end{tabular}
\end{table}

\section{Bound Motion and Perturbatively Consistent Resonance Geometry}
\label{sec:orbits}

Direct substitution of the additive Hamilton--Jacobi ansatz into Eq.~\eqref{eq:slow_metric} shows that the geodesic equation separates through first order in $\chi_\infty$; this is consistent with the separability structure of the slow rotation construction in Ref.~\cite{DattaSingha2026}. With
\begin{equation}
 S=-Et+L_z\phi+S_r(r)+S_\theta(\theta),
\end{equation}
and separation constant ${\cal K}=q^2$, we define
\begin{equation}
 \cos\iota=\frac{L_z}{q}.
\end{equation}
Using the Mino type parameter $d\tau=r^2d\lambda$ \cite{Mino2003,DrascoHughes2004,FujitaHikida2009},
\begin{align}
 \left(\frac{dr}{d\lambda}\right)^2&={\cal R}(r),
 \label{eq:r_motion}\\
 \left(\frac{d\theta}{d\lambda}\right)^2&=q^2-\frac{L_z^2}{\sin^2\theta},
 \label{eq:theta_motion}\\
 \frac{dt}{d\lambda}&=\frac{r^2}{A}(E-\omega L_z),
 \label{eq:t_motion}\\
 \frac{d\phi}{d\lambda}&=\frac{L_z}{\sin^2\theta}+\frac{r^2\omega E}{A},
 \label{eq:phi_motion}
\end{align}
with
\begin{equation}
 {\cal R}=r^2B\left[\frac{r^2}{A}(E^2-2\omega E L_z)-r^2-q^2\right].
 \label{eq:R}
\end{equation}
The polar frequency in $\lambda$ is $\Upsilon_\theta=q$.

For
\begin{equation}
 r_p=\frac{p}{1+e},\qquad r_a=\frac{p}{1-e},
\end{equation}
write
\begin{equation}
 E=E_0+\chi_\infty E_1,\qquad
 q=q_0+\chi_\infty q_1,
 \qquad
 \omega=\chi_\infty\bar\omega.
\end{equation}
The two turning point conditions give
\begin{equation}
 q_0^2=\frac{A_a-A_p}{A_p/r_p^2-A_a/r_a^2},
 \qquad
 E_0^2=A_p\left(1+\frac{q_0^2}{r_p^2}\right),
 \label{eq:q0E0}
\end{equation}
and
\begin{equation}
 q_1=\frac{E_0\cos\iota(\bar\omega_a-\bar\omega_p)}{A_p/r_p^2-A_a/r_a^2},
 \label{eq:q1_new}
\end{equation}
\begin{equation}
 E_1=\bar\omega_p q_0\cos\iota+\frac{A_pq_0q_1}{E_0r_p^2}.
 \label{eq:E1_new}
\end{equation}
Expanding ${\cal R}={\cal R}_0+\chi_\infty{\cal R}_1$ gives
\begin{align}
 {\cal R}_0&=r^2B\left(\frac{r^2E_0^2}{A}-r^2-q_0^2\right),
 \label{eq:R0}\\
 {\cal R}_1&=r^2B\left[\frac{r^2}{A}
 (2E_0E_1-2\bar\omega E_0q_0\cos\iota)-2q_0q_1\right].
 \label{eq:R1}
\end{align}

Perturbative consistency requires expanding the radial period itself rather than evaluating a finite $\chi_\infty$ square root that would retain uncontrolled higher order terms. We therefore use
\begin{align}
 \Lambda_r^{(0)}&=2\int_{r_p}^{r_a}\frac{dr}{\sqrt{{\cal R}_0}},
 \label{eq:L0}\\
 \Lambda_r^{(1)}&=-\int_{r_p}^{r_a}\frac{{\cal R}_1}{{\cal R}_0^{3/2}}\,dr,
 \label{eq:L1}
\end{align}
so that
\begin{equation}
 \Upsilon_r^{(0)}=\frac{2\pi}{\Lambda_r^{(0)}},\qquad
 \Upsilon_r^{(1)}=-\Upsilon_r^{(0)}\frac{\Lambda_r^{(1)}}{\Lambda_r^{(0)}}.
 \label{eq:Ur}
\end{equation}
Since coordinate time introduces the same factor $\Gamma=\langle dt/d\lambda\rangle$ in both frequencies,
\begin{equation}
 \frac{\Omega_\theta}{\Omega_r}=\frac{\Upsilon_\theta}{\Upsilon_r}.
 \label{eq:ratio_cancel}
\end{equation}
We locate the $3{:}2$, $5{:}3$, and $2{:}1$ roots of Eq.~\eqref{eq:ratio_cancel} for the prograde baseline orientation $L_z>0$. Unless stated otherwise, all three roots are evaluated at the same $(e,\iota)=(0.2,45^\circ)$ and therefore form a fixed slice through the conservative resonance surfaces, not a dissipative inspiral trajectory. When more than one commensurability root is present, we follow the outer librating branch continuously from the corresponding vacuum root; the reported benchmarks have simple radial turning points rather than a separatrix double root. The selected ratios are representative low order commensurabilities and are not assumed to generate the largest self force jump.

\section{Conservative Resonance Network}
\label{sec:network}

\subsection{Fixed global Dehnen sequence}

\begin{figure}[t]
\centering
\begin{minipage}[t]{0.32\linewidth}\centering
\includegraphics[width=\linewidth]{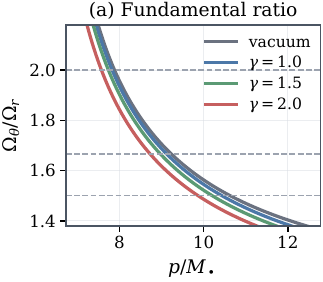}
\end{minipage}\hfill
\begin{minipage}[t]{0.32\linewidth}\centering
\includegraphics[width=\linewidth]{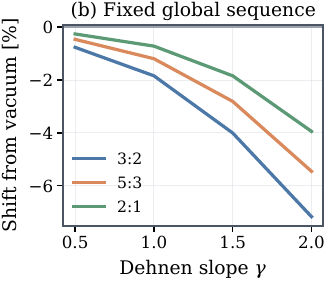}
\end{minipage}\hfill
\begin{minipage}[t]{0.32\linewidth}\centering
\includegraphics[width=\linewidth]{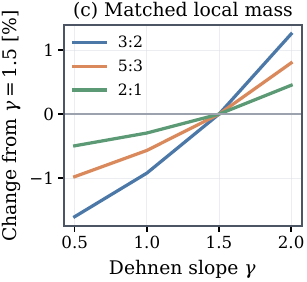}
\end{minipage}
\caption{Conservative resonance response. Panel (a) shows the polar to radial frequency ratio and the three selected commensurabilities. Panel (b) gives the resonance shifts as the Dehnen slope changes at fixed $(\mu,\alpha,\beta)=(1,100,2)$. Panel (c) instead fixes both $\mu=1$ and $\epsilon_{10}=0.01754248$; the residual nonparallel motion isolates radial profile information from these two mass normalizations.}
\label{fig:resonance}
\end{figure}

Table~\ref{tab:fixedroots} gives the strict first order resonance locations for the fixed global benchmark. The vacuum values are computed with the same numerical pipeline, including the rotational expansion. For $\gamma=2$, the resonance shifts relative to vacuum are $-7.170\%$, $-5.453\%$, and $-3.943\%$. Their unequal magnitude is the basic network effect: changing the radial matter distribution does not translate all commensurabilities by the same fractional amount. Figure~\ref{fig:resonance}(a,b) shows the corresponding frequency ratio curves and fixed global shifts.

\begin{table}[t]
\centering
\caption{Strict first order resonance locations $p_{\rm res}/M_\bullet$ for the fixed global Dehnen sequence.}
\label{tab:fixedroots}
\begin{tabular}{lccc}
\toprule
Model & $3{:}2$ & $5{:}3$ & $2{:}1$\\
\midrule
Vacuum & 10.628702 & 9.226871 & 7.881200\\
$\gamma=0.5$ & 10.547706 & 9.184447 & 7.861019\\
$\gamma=1.0$ & 10.433226 & 9.116890 & 7.824813\\
$\gamma=1.5$ & 10.203137 & 8.967770 & 7.736136\\
$\gamma=2.0$ & 9.866646 & 8.723690 & 7.570470\\
\bottomrule
\end{tabular}
\end{table}

\subsection{Matched total and local enclosed mass}

The fixed global sequence changes both radial structure and the enclosed mass sampled by the orbit. We therefore construct a second sequence satisfying
\begin{equation}
 \mu=1,\qquad \epsilon_{10}=0.01754248,
 \label{eq:matched_constraints}
\end{equation}
while $\gamma$ is varied and $\alpha$ is adjusted. The resulting models are listed in the upper part of Table~\ref{tab:matched}. Relative to the $\gamma=1.5$ model, the $3{:}2$ location changes by $-1.605\%$, $-0.923\%$, and $+1.254\%$ for $\gamma=0.5$, $1.0$, and $2.0$, whereas the corresponding $2{:}1$ changes are $-0.496\%$, $-0.295\%$, and $+0.452\%$. The response is therefore not eliminated by matching the total and local enclosed mass; Fig.~\ref{fig:resonance}(c) displays the resulting nonparallel displacement of the three commensurabilities.

For an amplitude independent representation of the ordered network, define
\begin{equation}
 {\cal S}_{53}=\frac{p_{5:3}}{p_{3:2}},\qquad
 {\cal S}_{21}=\frac{p_{2:1}}{p_{3:2}}.
 \label{eq:Scoords}
\end{equation}
These quantities describe the relative radial spacing of the three conservative commensurabilities.

\subsection{Orbital state dependence of the conservative surfaces}
\label{sec:orbitalstate}

\begin{figure}[t]
\centering
\includegraphics[width=0.98\linewidth]{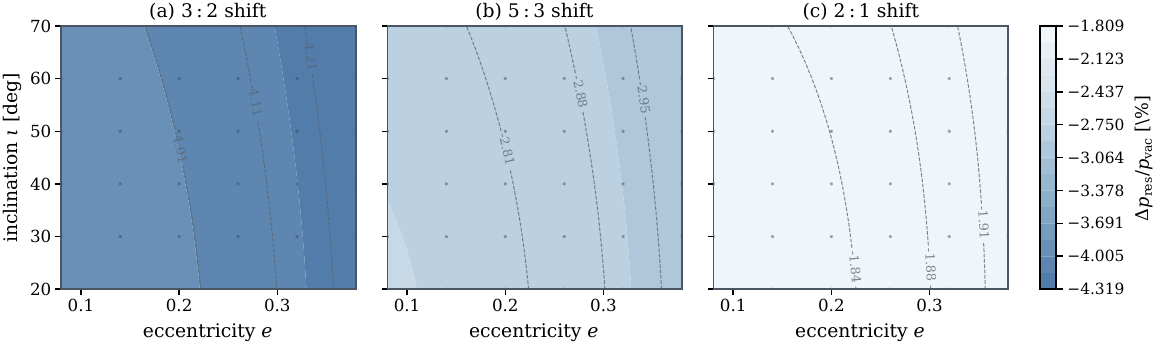}
\caption{Orbital state dependence of the conservative resonance displacement for the fixed global $\gamma=1.5$ Dehnen environment relative to vacuum. Panels (a)--(c) show the $3{:}2$, $5{:}3$, and $2{:}1$ fractional shifts over the domain in Eq.~\eqref{eq:orbital_state_domain}. The same color scale is used in all panels. The small points mark the $6\times6$ directly evaluated grid and the contours interpolate between those points for visualization only; all quantitative ranges quoted in the text are taken from the grid values themselves.}
\label{fig:orbitalstate}
\end{figure}

The benchmark slice at $e=0.2$ and $\iota=45^\circ$ is useful for controlled profile comparisons, but a single orbital state is not sufficient to establish that the differential resonance response survives ordinary changes in the bound orbit. We therefore evaluate the same fixed global $\gamma=1.5$ Dehnen environment over
\begin{equation}
 0.08\leq e\leq0.38,
 \qquad
 20^\circ\leq\iota\leq70^\circ,
 \label{eq:orbital_state_domain}
\end{equation}
with $\chi_\infty=0.05$ and $\eta=0.5$ held fixed. At each grid point we follow the outer librating branch separately for the environmental and vacuum geometries and form
\begin{equation}
 \Delta_a(e,\iota)
 =100\left[\frac{p_a^{\rm env}(e,\iota)}{p_a^{\rm vac}(e,\iota)}-1\right] .
 \label{eq:surface_shift}
\end{equation}
Figure~\ref{fig:orbitalstate} shows the three resulting resonance surface displacements. Across the directly evaluated grid, $\Delta_{3:2}$ ranges from $-4.32\%$ to $-3.90\%$, $\Delta_{5:3}$ from $-3.01\%$ to $-2.74\%$, and $\Delta_{2:1}$ from $-1.94\%$ to $-1.81\%$. Thus the ordering
\begin{equation}
 |\Delta_{3:2}|>|\Delta_{5:3}|>|\Delta_{2:1}|
 \label{eq:surface_ordering}
\end{equation}
holds at all 36 directly evaluated grid points rather than occurring only at the reference orbit. The absolute shifts vary smoothly with orbital state, so this test does not remove the nuisance degeneracies quantified below; it establishes the narrower point that the differential radial response of the network is not a single slice artifact.

\section{Cross Family Profile Test and Nuisance Structure}
\label{sec:crossfamily}

\subsection{Independent profile family}

The lower part of Table~\ref{tab:matched} repeats the matched test for the Einasto inspired family. The Einasto shape parameter $\alpha_E$ is varied and $r_{-2}$ is adjusted so that Eq.~\eqref{eq:matched_constraints} remains satisfied. The two families cover an overlapping range in $\kappa_{10}$ while differing in their global functional form.

\begin{table}[t]
\centering
\caption{Matched profile sequences with $\mu=1$ and $\epsilon_{10}=0.01754248$. The scale column is $\alpha=a_0/M_\bullet$ for Dehnen and the inherited untapered Einasto scale $r_{-2}/M_\bullet$ for the Einasto inspired family.}
\label{tab:matched}
\begin{tabular}{llrrrr}
\toprule
Family & Shape & Scale & $\kappa_{10}$ & ${\cal S}_{53}$ & ${\cal S}_{21}$\\
\midrule
Dehnen & $\gamma=0.5$ & 32.1537 & 2.40693 & 0.884514 & 0.766756\\
Dehnen & $\gamma=1.0$ & 50.4011 & 2.16888 & 0.882069 & 0.763012\\
Dehnen & $\gamma=1.5$ & 100.0000 & 1.86364 & 0.878923 & 0.758212\\
Dehnen & $\gamma=2.0$ & 354.8287 & 1.47259 & 0.874991 & 0.752202\\
\midrule
Einasto inspired & $\alpha_E=0.12$ & 11.8725 & 1.70358 & 0.876936 & 0.755297\\
Einasto inspired & $\alpha_E=0.16$ & 18.1020 & 1.85793 & 0.878541 & 0.757740\\
Einasto inspired & $\alpha_E=0.20$ & 22.5459 & 1.98690 & 0.879926 & 0.759833\\
Einasto inspired & $\alpha_E=0.30$ & 28.4509 & 2.23979 & 0.882735 & 0.764042\\
\bottomrule
\end{tabular}
\end{table}

\begin{figure}[t]
\centering
\begin{minipage}[t]{0.32\linewidth}\centering
\includegraphics[width=\linewidth]{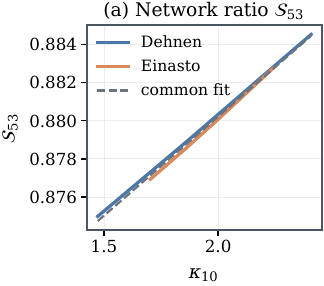}
\end{minipage}\hfill
\begin{minipage}[t]{0.32\linewidth}\centering
\includegraphics[width=\linewidth]{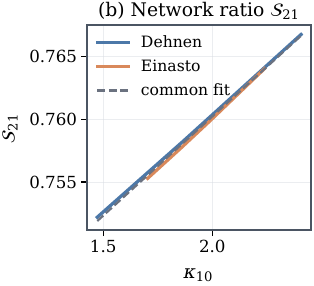}
\end{minipage}\hfill
\begin{minipage}[t]{0.32\linewidth}\centering
\includegraphics[width=\linewidth]{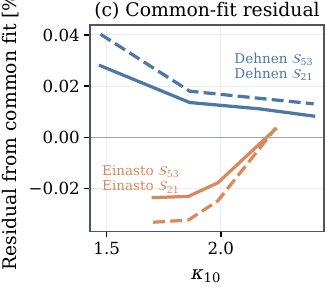}
\end{minipage}
\caption{Cross family local gradient test at fixed total environmental mass and fixed enclosed environmental mass at $10M_\bullet$. Panels (a) and (b) show that the two resonance spacing ratios follow the local logarithmic mass slope across the Dehnen and Einasto inspired families. Panel (c) shows the residual from a common linear fit; the small residual is a statement about the sampled domain, not a universal parametrization.}
\label{fig:crossfamily}
\end{figure}

Figure~\ref{fig:crossfamily} shows that the two network shape coordinates track $\kappa_{10}$ closely across both families. A common least squares linear fit over the sampled points gives
\begin{align}
 {\cal S}_{53}&=0.859467+0.010376\,\kappa_{10},
 \label{eq:fitS53}\\
 {\cal S}_{21}&=0.728642+0.015793\,\kappa_{10}.
 \label{eq:fitS21}
\end{align}
The largest fractional departures from these common fits are $2.8\times10^{-4}$ and $4.0\times10^{-4}$, respectively. Equations~\eqref{eq:fitS53} and \eqref{eq:fitS21} are empirical relations for the controlled domain in Table~\ref{tab:matched}, not universal halo laws. Their significance is that a response first identified inside one Dehnen family persists under a structurally different radial mass profile and is organized predominantly by a local mass gradient after the two mass normalizations are fixed.

The tight $\beta=2$ collapse does not by itself establish that $\kappa_{10}$ is a sufficient environmental coordinate, because both families use the same horizon regularization. We therefore repeat the matched Dehnen sequence with $\beta=3$ and $4$ and compare it with the common $\beta=2$ fits in Eqs.~\eqref{eq:fitS53} and \eqref{eq:fitS21}. Table~\ref{tab:taper} gives the maximum fractional departures. They increase systematically but remain at the few $10^{-3}$ level over the sampled domain. Thus $\kappa_{10}$ is a leading organizer of the network spacing, not a complete descriptor; higher radial structure and the strong field regularization remain measurable in principle.

\begin{table}[t]
\centering
\caption{Sensitivity of the matched network relation to the horizon taper. The $\beta=2$ row is the combined Dehnen--Einasto cross family test. The $\beta=3,4$ rows are matched Dehnen sequences compared with the common $\beta=2$ fits.}
\label{tab:taper}
\begin{tabular}{lcc}
\toprule
Profile test & $\max|\Delta {\cal S}_{53}/{\cal S}_{53}|$ & $\max|\Delta {\cal S}_{21}/{\cal S}_{21}|$\\
\midrule
$\beta=2$, Dehnen + Einasto inspired & $2.8\times10^{-4}$ & $4.0\times10^{-4}$\\
$\beta=3$, Dehnen & $7.3\times10^{-4}$ & $1.14\times10^{-3}$\\
$\beta=4$, Dehnen & $1.42\times10^{-3}$ & $2.09\times10^{-3}$\\
\bottomrule
\end{tabular}
\end{table}

A weak environment expansion clarifies why local radial structure can enter differently at distinct resonances. Write
\begin{equation}
 m(r)=M_\bullet+\varepsilon h(r),
 \qquad \varepsilon\ll1,
 \label{eq:weakm}
\end{equation}
with $h$ sufficiently horizon regular, for example $h={\cal O}[(r-2M_\bullet)^\beta]$ with $\beta\geq2$. To first order,
\begin{equation}
 \delta B(r)=-\frac{2h(r)}{r},
 \label{eq:dB}
\end{equation}
while Eq.~\eqref{eq:A_quad} gives
\begin{equation}
 \delta\ln A(r)=-2\int_r^\infty\frac{h(s)}{[s-2M_\bullet]^2}\,ds.
 \label{eq:dlnA}
\end{equation}
If ${\cal Q}(p;h)=\Omega_\theta/\Omega_r$ and $p_a^{(0)}$ is a vacuum root with ${\cal Q}=q_a$, then
\begin{equation}
 \delta p_a=-\left.\frac{\delta {\cal Q}[h]}{\partial_p{\cal Q}}\right|_{p_a^{(0)}}.
 \label{eq:rootshift}
\end{equation}
Different $p_a^{(0)}$ therefore evaluate different radial functionals of the same $h(r)$. This is the mathematical basis for a network response. It does not imply that three resonance radii uniquely reconstruct an arbitrary function $h(r)$.

\subsection{Degeneracy with orbital nuisance parameters}

To quantify the latter point, we augment the three resonance locations with conservative nodal precession,
\begin{equation}
 \Pi_a=\left.\frac{\Omega_\phi-\Omega_\theta}{\Omega_\theta}\right|_a.
 \label{eq:Pi}
\end{equation}
At first order in spin, spherical symmetry of the static problem implies
\begin{equation}
 \Pi_a=\chi_\infty\frac{E_0}{q_0}
 \left\langle\frac{r^2\bar\omega}{A}\right\rangle_{\lambda,r}
 +{\cal O}(\chi_\infty^2),
 \label{eq:Pi_first}
\end{equation}
where the radial average is evaluated on the zeroth order orbit. We form the six component observable vector
\begin{equation}
 \bm{o}=(\ln p_{3:2},\ln p_{5:3},\ln p_{2:1},
 \ln\Pi_{3:2},\ln\Pi_{5:3},\ln\Pi_{2:1})
\end{equation}
and the operational local response directions
\begin{equation}
 \bm{v}=(\kappa_{10},\ln\epsilon_{10},e,\iota,\ln\chi_\infty,\eta).
\end{equation}
The first entry is not an abstract variation of $\kappa_{10}$ at fixed higher profile moments: it is the directional derivative along the $\epsilon_{10}$ matched Dehnen sequence, using $\gamma=1.45$ and $1.55$ with $\alpha$ readjusted. The $\ln\epsilon_{10}$ column is the overall environmental amplitude direction obtained by varying $\ln\mu$ at fixed Dehnen shape. The remaining centered steps are $\Delta e=0.005$, $\Delta\iota=1^\circ$, $\Delta\ln\chi_\infty=0.02$, and $\Delta\eta=0.05$. With these operational definitions the finite difference response $J_{iA}=\partial o_i/\partial v_A$ is evaluated at the baseline model. Column normalization removes trivial parameter scaling before a singular value decomposition.

The normalized singular values are
\begin{equation}
 (2.2984,\,0.7801,\,0.2979,\,0.1413,\,0.00850,\,0.00295),
 \label{eq:singvals}
\end{equation}
corresponding to a condition number $\simeq7.8\times10^2$. Rescaling all finite difference steps by factors of $1/2$ and $2$ changes this value only to $780.36$ and $780.62$, respectively, so the poor conditioning is not a finite difference step artifact. This operational Jacobian is therefore numerically full rank but strongly ill conditioned. This distinction is essential: rank alone does not establish robust identifiability, and the result should not be interpreted as a global inverse theorem for the profile parameters.

\begin{figure}[t]
\centering
\includegraphics[width=0.99\linewidth]{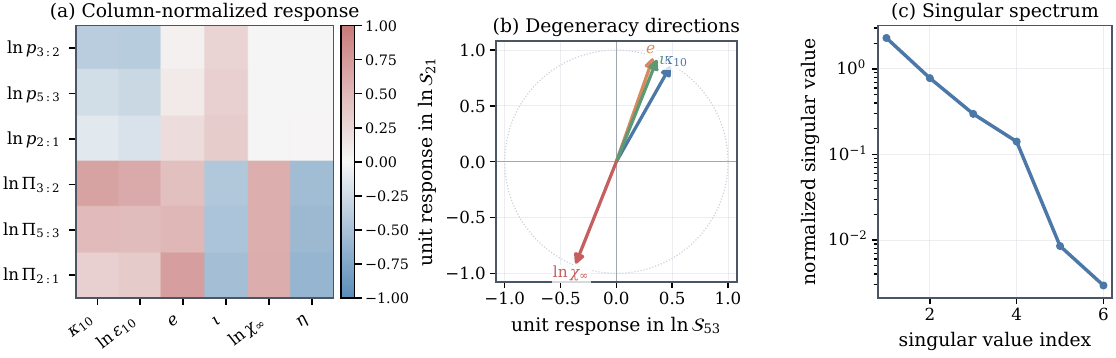}
\caption{Conservative nuisance structure at the baseline Dehnen model. Panel (a) shows the six observable response matrix after independent column normalization, exposing which observables respond similarly to different parameter directions. Panel (b) compares sign preserving unit normalized directions in the two network shape coordinates and therefore isolates degeneracy angles rather than derivative magnitudes. Panel (c) shows the singular spectrum of the same column normalized response matrix. The last two singular values are small despite numerical full rank, demonstrating the poor conditioning of a local conservative inversion.}
\label{fig:nuisance}
\end{figure}

The two dimensional network shape direction $(\partial\ln{\cal S}_{53},\partial\ln{\cal S}_{21})$ is also nearly aligned or anti-aligned with ordinary orbital variations. Its cosine with the eccentricity response is $0.9847$, while the absolute cosines with inclination and spin are $0.9908$. The corresponding acute angles are only about $10.0^\circ$ and $7.8^\circ$. Figure~\ref{fig:nuisance} visualizes both the alignment and the singular spectrum. Conservative resonance spacing is therefore sensitive to local environmental structure but cannot by itself be interpreted as a stable profile inversion when the orbital parameters are allowed to vary.

\section{Connection to Dissipative Inspiral Observables}
\label{sec:scope}

The conservative result above is intentionally separated from inspiral timing. This separation is required by the current error budget.

First, the Peters--Mathews fluxes are derived in the weak field, slow motion limit \cite{PetersMathews1963,Peters1964}. Strong field EMRI evolution generally requires relativistic flux information; improved approximate Kerr schemes were developed to extend weak field transport into the relativistic regime \cite{GairGlampedakis2006}, and a relativistic perturbative framework for radiation in black hole matter environments has recently been formulated \cite{PolcarWitzany2025}. Recent matter environment calculations find that quadrupole fluxes can agree well with Teukolsky results over broad configurations, but they also compare post Newtonian and Teukolsky based prescriptions explicitly when modeling resonance crossings \cite{StratenyEtAl2026}. No corresponding subpercent error bound has been established for Peters transport on the environmental backgrounds used here. We therefore do not attach an inference precision to timing observables based on such a prescription.

Second, an inspiral does not cross successive resonances smoothly. Resonant self force harmonics generate changes in the orbital actions whose sign and amplitude depend on the resonant phase and forcing coefficients \cite{FlanaganHinderer2012,FlanaganHughesRuangsri2014,LevatiCardenas2026}. Such a jump changes the orbit entering the next crossing and therefore propagates directly into any inter resonance clock or accumulated phase ratio. Those observables cannot be used consistently here because the environmental resonant self force coefficients are not available.

Third, matter induced dissipation can alter the inspiral rate. Dynamical friction and accretion have measurable consequences in several dark matter EMRI models \cite{HannukselaEtAl2020,ZhangFuDai2024,DuqueEtAl2024}. The effective stress tensor in Eq.~\eqref{eq:stress} specifies the spacetime source but not the phase space distribution, Coulomb logarithm, capture prescription, or microphysics required for a unique drag force. Assigning a numerical drag correction from the metric alone would therefore add an uncontrolled model assumption.

For these reasons, timing ratios, accumulated phase ratios, resonance duration ratios, and detector precision forecasts are not part of the claimed result. A future dissipative extension should combine a relativistic environmental flux model with phase dependent resonance jumps and environment specific drag before these observables are used for inference.

\section{Numerical Verification and Limiting Cases}
\label{sec:numerics}

The background equations are integrated in $u=\ln r$ from an outer boundary $r_{\max}=4\times10^5M_\bullet$. The asymptotic conditions are $A=1-2M_{\rm ADM}/r+{\cal O}(r^{-2})$ and $\bar\omega=2M_\bullet^2/r^3+{\cal O}(r^{-4})$. Radial period integrals are regularized with
\begin{equation}
 r(\psi)=\frac{p}{1+e\cos\psi},\qquad 0\leq\psi\leq\pi,
 \label{eq:anomaly}
\end{equation}
and evaluated by Gaussian quadrature. Resonance roots are bracketed directly. If more than one sign change is present, the largest $p$ root connected continuously to the vacuum outer librating branch is selected; this branch prescription avoids spuriously switching to an inner near separatrix root as eccentricity increases.

For the baseline $\gamma=1.5$ model, increasing the quadrature order from 256 to 384 changes all three roots by at most $2.7\times10^{-9}$ in relative terms. Increasing the outer boundary from $4\times10^5M_\bullet$ to $8\times10^5M_\bullet$ changes them by at most $8.7\times10^{-10}$, while tightening the background integration tolerance from $2\times10^{-10}$ to $2\times10^{-11}$ changes them by at most $1.2\times10^{-9}$. These numbers characterize internal numerical convergence only; they do not estimate omitted physical effects.

The weak environment expansion in Sec.~\ref{sec:crossfamily} is also checked directly. At fixed $(\alpha,\gamma,\beta)=(100,1.5,2)$ and $\mu=(0.01,0.03,0.10)$, the three resonance shifts remain ordered and approximately linear in $\mu$: across these amplitudes the shift divided by $\mu$ varies by only $1.36\%$, $0.98\%$, and $0.66\%$ for the $3{:}2$, $5{:}3$, and $2{:}1$ roots, respectively. This test supports the perturbative interpretation of the network response in the dilute limit without turning the $\mu=1$ benchmark into an astrophysical population prior.

A stronger analytic check is available in Schwarzschild. In the circular limit,
\begin{equation}
 \frac{\Omega_\theta}{\Omega_r}=\frac{1}{\sqrt{1-6M_\bullet/p}},
 \label{eq:schw_circular_ratio}
\end{equation}
which gives $p/M_\bullet=(10.8,9.375,8)$ for the $(3{:}2,5{:}3,2{:}1)$ resonances. Evaluating the eccentric code at $e=0.01$ and $\chi_\infty=0$ gives $(10.800057,9.375069,8.000101)$, with a maximum relative discrepancy $1.3\times10^{-5}$ from the circular result. At $e=0.2$ and $\chi_\infty=0$, the roots are unchanged to numerical precision when the inclination is varied from $20^\circ$ to $70^\circ$, as required by spherical symmetry.

Several analytical limits provide independent checks. In the limit $\mu\rightarrow0$, Eqs.~\eqref{eq:mass_eq}--\eqref{eq:A_quad} reduce to Schwarzschild and Eq.~\eqref{eq:omega} reduces to the vacuum $r^{-3}$ frame dragging solution. In the limit $\chi_\infty\rightarrow0$, the resonance locations become independent of inclination because the static background is spherically symmetric. The parameter $\eta$ enters only the rotational sector and therefore disappears from the static resonance geometry. Finally, $\epsilon_{10}\rightarrow0$ continuously recovers the vacuum resonance network.

The dominant energy condition $p_t\leq\rho$ is equivalent in this $p_r=0$ model to $5m(r)\leq2r$. It is satisfied over the bound orbital intervals used for the reported benchmarks but, as already stated, is not imposed arbitrarily close to the horizon. In the two dimensional orbital state test the smallest sampled environmental pericenter is $r_p=5.63M_\bullet$, so none of the reported resonance surfaces directly probes the near horizon layer where the effective matter model becomes least physical. This distinction prevents an effective strong field environment from being misidentified as a globally viable microscopic dark matter distribution.

\section{Conclusion}
\label{sec:conclusion}

We have isolated the conservative information carried by fixed $(e,\iota)$ slices of several low order polar to radial resonance surfaces in source consistent black hole environments. A fixed global Dehnen sequence produces nonuniform shifts of the $3{:}2$, $5{:}3$, and $2{:}1$ commensurabilities. After matching both the total environmental mass and the enclosed mass at $10M_\bullet$, a residual multi resonance response remains. A Dehnen--Einasto comparison shows that the two spacing ratios are organized primarily by the local logarithmic enclosed mass slope, while the taper variation demonstrates that this slope is not a sufficient one parameter description of the strong field environment.

The physically defensible conclusion is therefore a local conservative diagnostic rather than unique tomography. The profile direction is nearly degenerate, up to orientation, with eccentricity, inclination, and rotation responses, and the six observable response matrix is strongly ill conditioned even under finite difference step variations. The independent weak amplitude check recovers the expected near linear environmental scaling for $\mu\leq0.1$, while the $\mu=1$ sequence remains a controlled strong field stress test rather than a population model. A sequential inspiral network would additionally require the evolution of the orbital elements, relativistic environmental fluxes, phase dependent resonance jumps, and a specified drag model. The present construction identifies a reproducible conservative profile sensitivity and delineates the additional physics required before it can be promoted to waveform level environmental inference.

\end{document}